\documentclass{article}

\usepackage{arxiv}

\usepackage[utf8]{inputenc} 
\usepackage[T1]{fontenc}    
\usepackage{hyperref}       
\usepackage{url}            
\usepackage{booktabs}       
\usepackage{amsfonts}       
\usepackage{nicefrac}       
\usepackage{microtype}      
\usepackage{lipsum}		
\usepackage{graphicx}
\usepackage{doi}
\usepackage{color}
\usepackage{tabularray}
\usepackage{tabularx}

\title{Data Cooperatives: Democratic Models for Ethical Data Stewardship}

\author{ \href{https://orcid.org/0009-0003-4478-138X}{\includegraphics[scale=0.06]{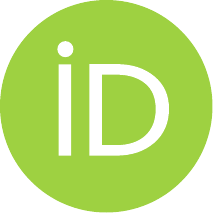}\hspace{1mm}Francisco Mendonça} \\
        University of Geneva\\
	  Geneva, Switzerland \\
	\texttt{francisco.de-andrade@etu.unige.ch} \\
	\And
	\href{https://orcid.org/0000-0001-5048-5251}{\includegraphics[scale=0.06]{orcid.pdf}\hspace{1mm}Giovanna Di Marzo} \\
        University of Geneva\\
	  Geneva, Switzerland \\
	\texttt{Giovanna.DiMarzo@unige.ch} \\
	\And
	\href{https://orcid.org/0000-0003-4554-6958}{\includegraphics[scale=0.06]{orcid.pdf}\hspace{1mm}Nabil Abdennadher} \\
	University of Applied Sciences of Western Switzerland \\
	  Geneva, Switzerland \\
	\texttt{nabil.abdennadher@hesge.ch} \\
}

\renewcommand{\shorttitle}{\textit{Data Cooperatives: Democratic Models for Ethical Data Stewardship}}

\hypersetup{
pdftitle={Data Cooperatives: Democratic Models for Ethical Data Stewardship},
pdfauthor={Francisco Mendonça},
pdfkeywords={Data Cooperatives, Data Management, Cooperatives},
}

\begin{document}
\maketitle

\begin{abstract}

Data cooperatives offer a new model for fair data governance, enabling individuals to collectively control, manage, and benefit from their information while adhering to cooperative principles such as democratic member control, economic participation, and community concern. This paper reviews data cooperatives, distinguishing them from models like data trusts, data commons, and data unions, and defines them based on member ownership, democratic governance, and data sovereignty. It explores applications in sectors like healthcare, agriculture, and construction. Despite their potential, data cooperatives face challenges in coordination, scalability, and member engagement, requiring innovative governance strategies, robust technical systems, and mechanisms to align member interests with cooperative goals. The paper concludes by advocating for data cooperatives as a sustainable, democratic, and ethical model for the future data economy.
\end{abstract}


\section{Introduction}
\label{sec:intr}
The increased harvesting and utilization of personal data by large corporations has become a signature of the digital economy. The trend has provoked profound concerns regarding data privacy, transparency, and exploitation and has engendered a call for alternative models of data governance. Data cooperatives have emerged as a promising solution, offering a model of collective data management and benefit by the people. Founded upon the proven principles of cooperative enterprises \cite{ica}, data cooperatives allow members to pool their data together, have democratic oversight over its use, and share in the value created. This is in contrast to traditional data management, in which individuals have little control and their data is commoditized.

Traditional cooperatives have a long history of success across a variety of industries, including agriculture \cite{farmingUK}, retail \cite{europarl}, and finance \cite{ilo}, demonstrating the potential of collective action for economic and social empowerment.  These organizations operate according to a set of core principles, including voluntary membership, democratic member control, economic participation, autonomy, education, cooperation, and concern for the community \cite{Cheney2023-oq}.  Data cooperatives seek to apply the same principles to the governance of data, in recognition of its increasing economic and social value \cite{Hardjono2019-gy}.  By pooling their data, members are able to bargain collectively with data consumers, protect their privacy, and ensure equitable distribution of benefits \cite{Girish2023-oe}.

The emergence of data cooperatives is a direct response to the concentration of power in the hands of large corporations, who are likely to behave with limited transparency and accountability in the use of data \cite{Buhler2023-vf}.  Data cooperatives present a potential means to correct this power balance, enabling individuals to have greater control over their own data and to participate in the data economy on more level terms \cite{Buhler2023-vf}.  However, applying cooperative principles to the digital environment poses specific challenges.  These include technical problems with respect to the control of data, ethical concerns with respect to the use of data, and the need for appropriate regulatory mechanisms \cite{Eke2024-jz}.

One of the most significant characteristics of data cooperatives is their fiduciary duty to members \cite{noauthor_undated-oh}. This implies a duty to act in the best interests of the members, to prioritize their needs, and to be transparent and accountable in all data activities \cite{Patel2021-js}. This includes robust data protection practices, ethical data use practices, and empowering members through data sovereignty – control over how their data is used \cite{noauthor_undated-bm}.

This paper offers a thorough analysis of data cooperatives, covering their definition, frameworks, applications, and challenges. 
The following sections are organized as follows: Section 2 defines data cooperatives and distinguishes them from related concepts like data commons, data trusts, and data unions.  Section 3 describes the four key frameworks that underpin data cooperatives: governance, operational, technical, and social.  Section 4 examines the diverse applications of data cooperatives across various industries, highlighting real-world examples and use cases.  Section 5 delves into the challenges associated with establishing and scaling data cooperatives, focusing on coordination, scalability, and member engagement.  Section 6 explores potential avenues for future research in the field of data cooperatives.  Finally, Section 7 concludes the paper with a summary of key findings and a discussion of the potential of data cooperatives to shape a more equitable and participatory data economy.

\section{Definition of Data Cooperatives}
\label{sec:dataCoopDef}

The debate around data governance is often muddled by the blurring of concepts like \textit{data commons}, \textit{data trusts}, and \textit{data unions} with \textit{data cooperatives}. While these models share commonalities, their distinctions need to be accurately delineated. This section explains the concept of data cooperatives, highlighting their unique characteristics and marking their boundaries from related concepts.

\subsection{Data Cooperatives}
\label{dataCoop}
We define a data cooperative as a member-owned organization where individuals voluntarily contribute their data, control its usage collectively, and share the resulting benefits fairly, with each individual retaining ownership of their personal data \cite{Hardjono2019-gy}. Imperatively, we hold the view that an organization can be called a data cooperative only if it adheres to and actively advocates for the seven basic cooperative principles: voluntary membership, democratic member control, economic participation, autonomy, education, cooperation, and concern for the community. These cooperative principles, upon which the cooperative business model is founded, form an ethical basis of data stewardship that prioritizes collective well-being over profit maximization, which differentiates data cooperatives from traditional data platforms.

We deliberately use the term "Data Stewardship" rather than "Data Ownership" to describe the cooperative's relationship with member-contributed data. While members retain fundamental rights regarding their own data, the cooperative itself doesn't assume traditional ownership of the combined data pool. "Ownership", in its conventional sense, implies exclusive control and the right to exclude others, which goes against the collaborative nature of a data cooperative and the non-rivalrous nature of data itself. Instead, "Stewardship" emphasizes the cooperative's responsibility to manage, protect, and facilitate the shared use of data for the collective benefit of its members. The cooperative acts as a trusted manager, enforcing agreed-upon rules and policies, ensuring data safety, accessibility, and reliability, on behalf of its members, who collectively govern its operations. The cooperative does not own the data; it stewards it.

\subsection{Data Commons}
\label{dataCommons}

A data commons is a model in which a community collaboratively controls, accesses, and utilizes data, under shared terms agreed in a manner conducive to fair access and collaboration \cite{Grossman2016-ln}.  Because both data commons and data cooperatives both prioritize community benefit, one key difference between them is in terms of locus of control.  In a data commons, data is a shared asset, and terms of governance often determined through a community and not individual owners of the information \cite{openfuture}.  Consequently, individual data sovereignty is less emphasized in a data commons compared with a cooperative model.

An example of data commons is the \textbf{OCC NOAA Data Commons} \cite{occdataEnvironmentalData}, a partnership between NOAA (National Oceanic and Atmospheric Administration), the OCC (Open Commons Consortium), Amazon, Google, IBM and Microsoft.  This data commons pools large amounts of data from NOAA's environmental resources, making it openly accessible to researchers, business, and the public.  

\subsection{Data Trusts}
\label{dataTrust}

A Data Trust is a legal framework where trustees are appointed to manage and govern data on behalf of individuals or groups, ensuring that the data is used according to agreed-upon terms and in the contributors’ best interests \cite{Chan2023-ad, Delacroix2019-up}. In contrast, Data Cooperatives are member-driven organizations where individuals pool their data to collectively govern its use, aiming to create value for their members through shared control and democratic decision-making. 

The Mayo Clinic, one of the largest medical institutions in the United States, established a Data Trust to manage sensitive patient data from various sources, including clinical care, education, and research \cite{Chute_Beck_Fisk_Mohr_2010}. This trust operates under strict governance rules, which ensures a high level of security.

\subsection{Data Unions}
\label{dataUnion}

A data union is an organizational setup where individuals join together and negotiate for improved terms—e.g., monetary compensation, greater privacy protections, or data portability—when giving data to a third party.  Its ultimate goal is for its members to receive a direct advantage based on the value of their data \cite{californialawreviewDataUnions}.  Data unions most commonly leverage its members' bargaining power to negotiate with a third party, e.g., corporation or platform, regarding terms like equitable compensation for use of its data or better privacy terms.

The key differentiation between data unions and data cooperatives is in orientation.  Data unions have an orientation towards external negotiation in an attempt to secure beneficial terms for shared data, while data cooperatives have an orientation towards governance, in that members together have governance, use, and benefit over shared data *within* a cooperative.  Where data unions have an interest in maximising value in shared data when shared with an external party, data cooperatives have an interest in fair governance and use of data *among* its members.  This orientation creates significant differentiation in governance model, operational model, and form of member participation.

\subsection{Comparative Table}
\label{compTable}
Table \ref{tab:dataModels} presents a comparative analysis of data cooperatives, data unions, data commons, and data trusts, highlighting key differences in their characteristics. The table summarizes the most significant factors - ownership, governance, focus, value creation, data sovereignty, membership, and purpose - to facilitate a clear comparison of these models.

\textit{Ownership} refers to the legal and practical ownership of data, whether it's held by individuals, the collective, trustees, or the community. \textit{Governance} encompasses the decision-making processes and structures, which include member control, collective bargaining, or trustee management. \textit{Focus} indicates whether the model prioritizes internal governance, external negotiations, or community-driven goals. \textit{Value Creation} describes how value is generated and distributed to the members, be it financial returns, access to improved analytics, or other shared benefits. \textit{Data Sovereignty} covers the level of individual control over the data and how it's used. \textit{Membership} outlines the criteria and nature of membership, be it voluntary, open, or defined by shared interests. Finally, \textit{Purpose}, describes the overarching goals of the model. 

\begin{table}[!htbp]
\centering
\begin{tabularx}{\textwidth}{lXXXXX} 
\toprule
Aspect             & Data Cooperatives & Data Unions & Data Commons & Data Trusts \\
\midrule
\textbf{Ownership}        & Member-owned, collective control & Pooled for negotiation, individual ownership varies & Community-managed, shared resource & Trustees hold legal ownership, manage on behalf of beneficiaries \\
\midrule 

\textbf{Governance}       & Democratic, member-driven, one member one vote (typically) & Collective bargaining, member mandates & Shared governance, community rules & Trustee-managed, guided by trust deed \\
\midrule 

\textbf{Focus}            & Internal governance, equitable data use among members & External negotiations for better data terms & Equitable access, collaborative data utilization & Responsible data management, beneficiary interests \\
\midrule 

\textbf{Value Creation}   & Data products/services, collective insights, potential financial returns & Improved terms with third parties (e.g., compensation, privacy) & Shared resource access, community benefits & Benefits as defined in trust terms, often focused on stewardship \\
\midrule 

\textbf{Data Sovereignty} & Members retain control over their data, decide usage & Members retain some control, focus on external terms & Less emphasis on individual sovereignty, community-centric & Individuals relinquish some control for legal protection, trust-based management \\
\midrule 
\textbf{Membership}       & Voluntary, defined criteria & Voluntary, often focused on shared interests/needs & Open to community, defined criteria may exist & Beneficiaries defined in trust deed, may be individuals or groups \\
\midrule 

\textbf{Purpose}          & Collective benefit, empowerment, ethical data use & Maximizing individual returns from data sharing & Advancing shared knowledge, research, or other community goals & Protecting data rights, ensuring responsible use, often for a specific purpose \\

\bottomrule
\end{tabularx}
\caption{Comparative Analysis of Data Governance Models}
\label{tab:dataModels}
\end{table}

\newpage
\section{Data Cooperative Frameworks}
\label{sec:dataCoopFrame}

A data cooperative, being a sophisticated social technology, can intellectually be organized in terms of four interrelated pillars, each one a critical aspect of its function and role.  Governance, operational, technical, and social, these four pillars represent an overall model for studying and describing the multi-faceted nature of data cooperatives.

\subsection{Governance Framework}

\begin{figure}[!htbp]
\centering
\includegraphics[width=0.6\textwidth]{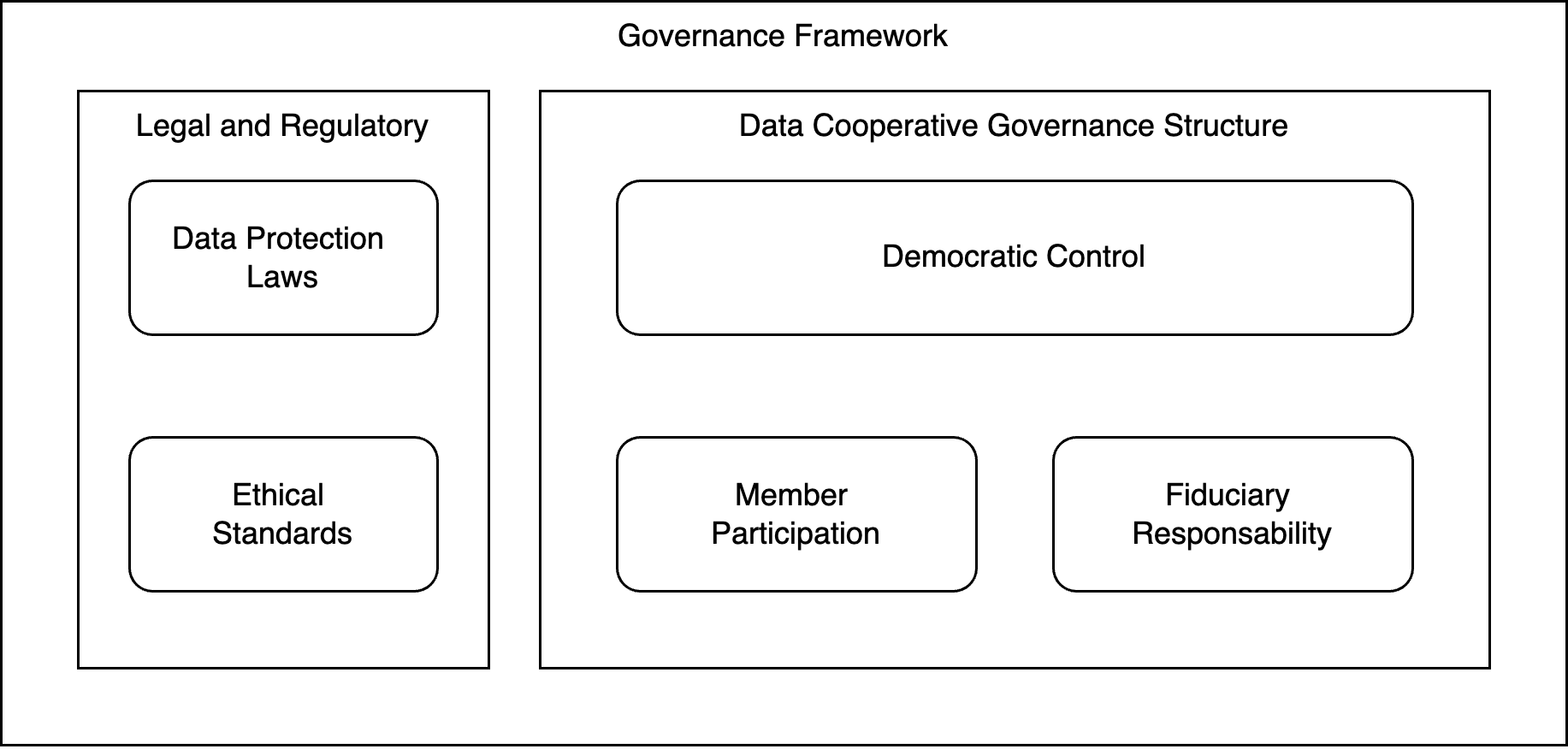}
\caption{Core principles of governance in data cooperatives.}
\label{fig:governance-framework}
\end{figure}

The governance model (Figure \ref{fig:governance-framework}) forms the backbone of a data cooperative, outlining decision-making processes and ensuring its legality and ethics.  Two closely intertwined components in this pillar include compliance with legal and regulative frameworks and the cooperative governance model.

The governance model of the cooperative is premised on the principle of democratic member control \cite{Hardjono2019-gy}.  This necessitates establishment of concrete structures for member participation in decision-making, such that members have an effective voice in deciding the direction and policies of the cooperative.  Democratic control is most often facilitated through such structures as voting, board governance, and open forums for discussion and feedback.  Governance must, in addition, specify the cooperative's bylaws and establish fiduciary requirements for leadership in the cooperative, such that member interests are protected and a platform for ethical operations is established \cite{Buhler2023-vf}.

Supplementing the inner governance structure is compliance with laws and regulations.  Data cooperatives must operate in terms of applicable laws and legislation, such as data protection legislation such as GDPR, and comply with key ethical values such as equity, transparency, and accountability \cite{noauthor_undated-oh}.  Compliance with laws and legislation is not a formality but a critical consideration in developing trust between and amongst its members and in assuring long-term viability for the cooperative.  The cooperative's governance structure and compliance with laws and legislation together assure that the data cooperative is member-owned, ethics-based, and legally compliant. This pillar forms the base in terms of which the other pillars operate.

\subsection{Operational Framework}

\begin{figure}[!htbp]
\centering
\includegraphics[width=0.6\textwidth]{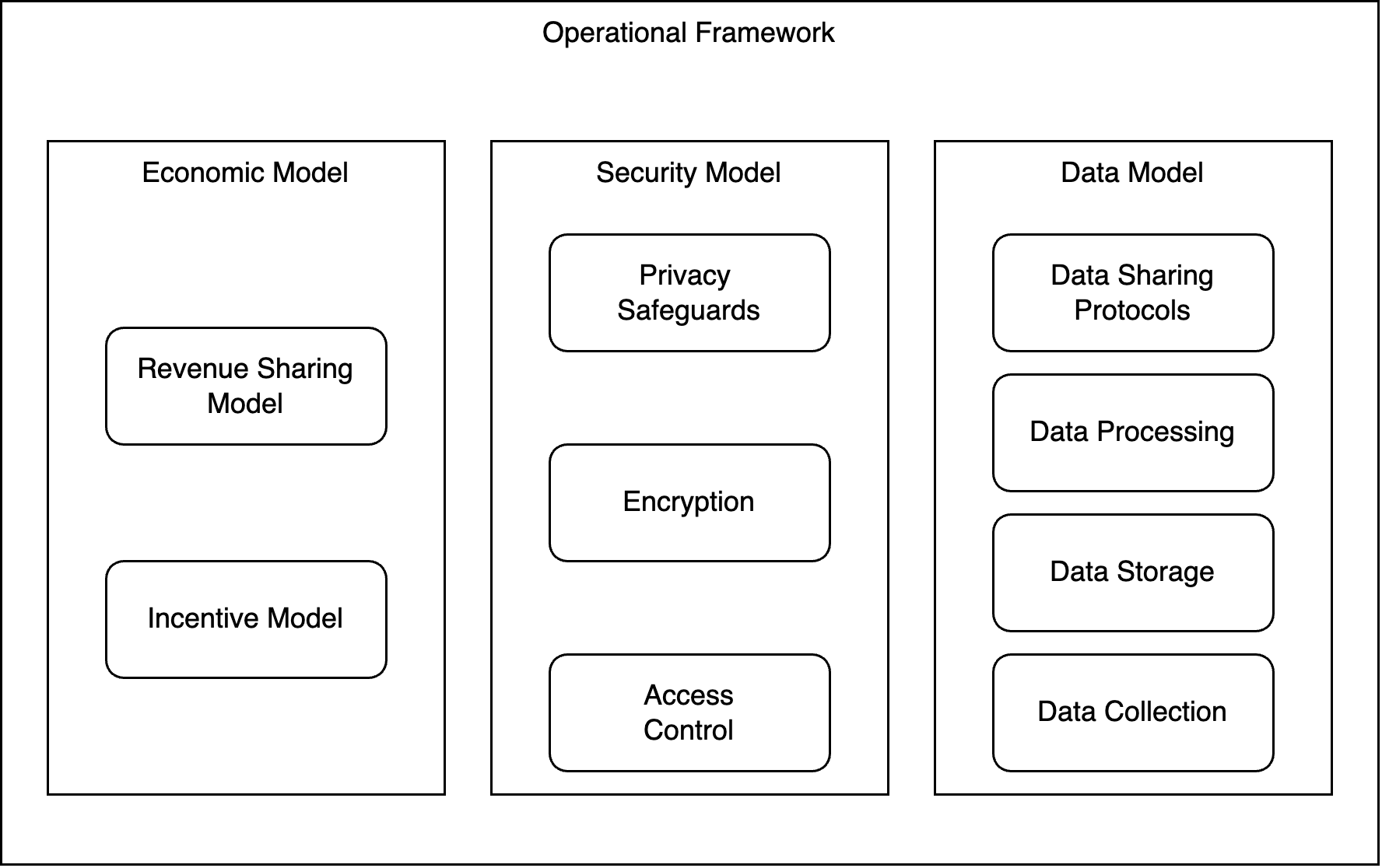}
\caption{Operational models for economy, data handling, and security.}
\label{fig:operational-framework}
\end{figure}

The operational framework (Figure \ref{fig:operational-framework}) serves as the bridge between governance principles and tangible outcomes.  It details the mechanics of how a data cooperative functions, translating overarching goals into concrete processes for managing data, safeguarding privacy, and generating value for its members. This framework is not a monolithic structure but rather a carefully interwoven system of three key layers: data management and usage protocols, privacy and security safeguards, and the economic and incentive architecture.

The foundation of this operational pillar lies in the meticulously crafted data management and usage protocols.  These protocols define the entire lifecycle of data within the cooperative, from the initial collection and secure storage to processing, analysis, and controlled sharing.  They are the practical embodiment of the cooperative's commitment to responsible data stewardship, ensuring transparency and accountability in all data-related activities \cite{Buhler2023-vf}.  These protocols are not static documents but rather living guidelines, subject to regular review and adaptation in response to evolving best practices, technological advancements, and member feedback.  They address critical aspects such as data quality control, metadata management, and data interoperability, laying the groundwork for effective data utilization.

Protecting member data is paramount, and the privacy and security safeguards layer addresses this critical concern.  Robust measures are implemented to ensure the confidentiality, integrity, and availability of personal information, mitigating the risks of unauthorized access, misuse, or breaches \cite{Buhler2023-wa}.  This layer often involves a multi-faceted approach, combining technical solutions like encryption and anonymization with organizational policies such as access control lists and data governance committees.  A key element is the establishment of clear roles and responsibilities, defining who has access to what data and under what circumstances.  Audit trails play a vital role, meticulously logging data access and modifications to maintain a transparent and accountable environment.  Furthermore, a dedicated data governance committee, elected by the members, provides ongoing oversight of information policies, regularly reviews security protocols, and acts swiftly and decisively in the event of any security incident, reinforcing member trust and ensuring compliance with the ever-changing landscape of data protection legislation.

Finally, the economic and incentive architecture defines how the cooperative generates and distributes value among its members.  This architecture is not simply about financial returns but encompasses a broader range of benefits, which may include access to data-driven insights, enhanced negotiating power in the marketplace, or preferential access to services offered by the cooperative \cite{mehta2022}. The challenge lies in designing a model that is both sustainable and equitable, incentivizing member participation and data sharing while ensuring that the benefits are distributed fairly and transparently.  This often involves striking a delicate balance between individual contributions and collective gains, fostering a sense of shared ownership and purpose.

These three pillars—data management protocols, privacy and security safeguards, and the economic and incentive architecture—are not isolated entities but rather integral components of a cohesive operational framework.  They work in synergy to translate the cooperative's governance principles into tangible actions, ensuring responsible data handling, robust privacy protections, and the creation of sustainable value for the cooperative's members.

\subsection{Technological Framework}

\begin{figure}[!h]
\centering
\includegraphics[width=0.6\textwidth]{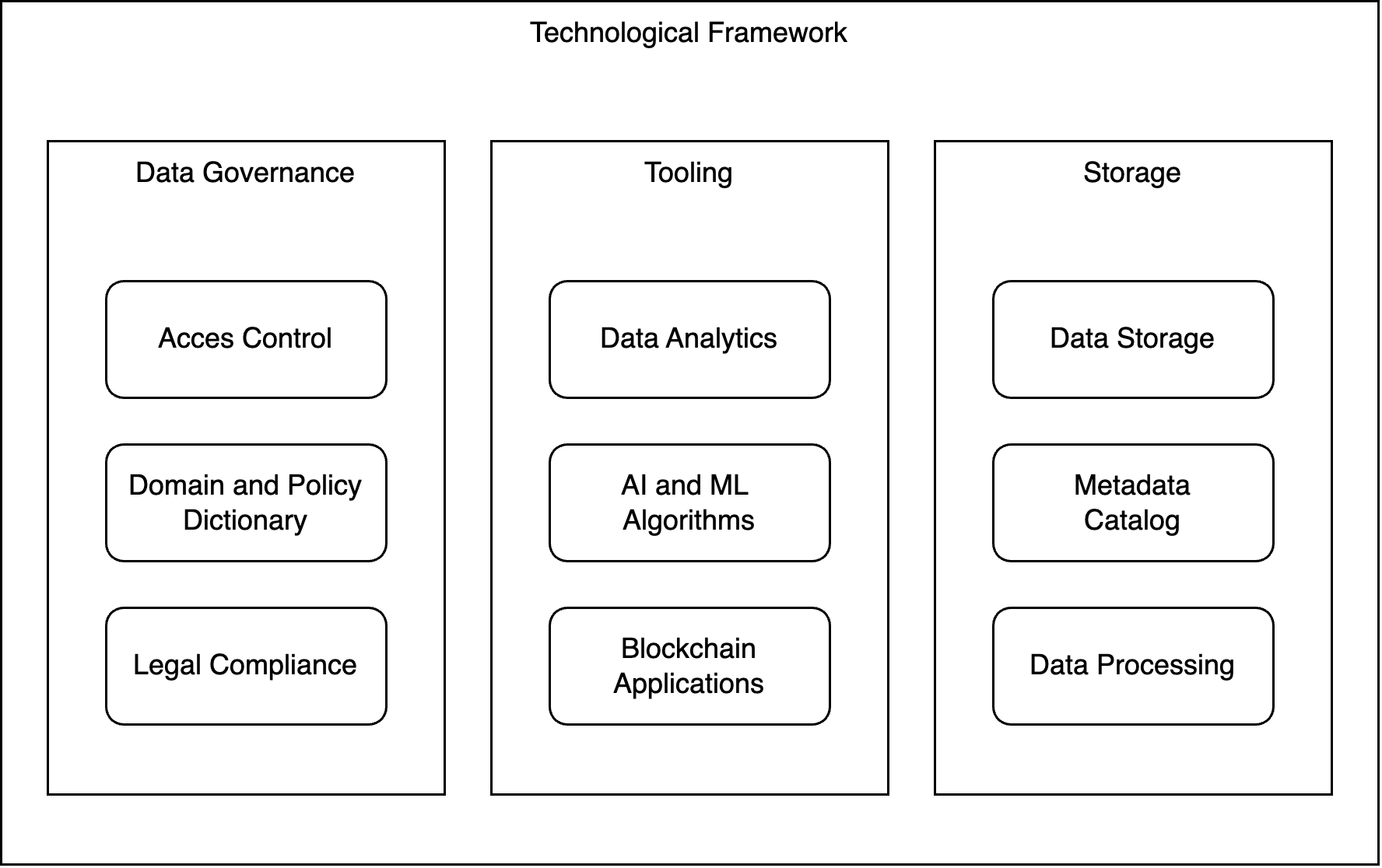}
\caption{Technological foundation for Data Cooperatives.}
\label{fig:tech-framework}
\end{figure}

The technological framework (Figure \ref{fig:tech-framework}) underpins the entire data cooperative ecosystem, providing the essential infrastructure for data storage, processing, digital security, interoperability, and scalability.  It is the tangible manifestation of the cooperative's commitment to effective and responsible data management, enabling both day-to-day operations and the pursuit of innovative data-driven initiatives.

A central element of this framework is the data governance platform.  This platform, often a sophisticated software system, serves as the control center for data access, usage monitoring, and compliance enforcement \cite{Hafen2019-jx}.  It is the mechanism through which the cooperative's data governance policies, established by the members, are put into practice.  The platform must be capable of managing granular access rights, ensuring that data is only accessible to authorized individuals or entities for approved purposes.  It also provides the tools for monitoring data usage, detecting anomalies, and generating audit trails, which are essential for maintaining transparency and accountability.  Furthermore, the platform plays a crucial role in ensuring compliance with a complex web of legal and ethical standards, automating compliance checks and providing alerts when potential violations are detected.

The technological stack supporting the data cooperative must be robust and versatile enough to handle a diverse range of data-related tasks.  It needs to provide the computational power and storage capacity for data analytics, enabling the cooperative to extract valuable insights from its collective data assets.  Furthermore, the stack should be adaptable to the integration of artificial intelligence (AI) and machine learning (ML) capabilities, allowing the cooperative to leverage these technologies for tasks such as predictive modeling, pattern recognition, and personalized services.  Scalability is a critical consideration, as the technological infrastructure must be able to accommodate the growing volume of data and increasing number of members as the cooperative expands.  Security is paramount, requiring robust measures to protect against cyber threats, data breaches, and unauthorized access.

While the benefits of data analytics and AI/ML are well-established, the role of blockchain technology within the data cooperative framework is still evolving, though its potential is becoming increasingly clear.  Beyond its well-known applications in cryptocurrencies, blockchain can offer unique advantages in enhancing transparency and trust within the cooperative.  By securely recording transactions and data-related activities on a distributed ledger, blockchain enables the creation of verifiable and auditable activity logs.  This can be particularly valuable in demonstrating compliance with data usage policies, ensuring accountability, and reinforcing member trust.  Furthermore, blockchain can facilitate secure data sharing and collaboration, enabling members to share data with external partners in a transparent and controlled manner, aligning with the cooperative’s principles of data sovereignty and member control.

The seamless integration of these technological elements—data governance platforms, robust technological stacks, and emerging technologies like blockchain—is essential for the successful operation of a data cooperative.  This integration ensures efficient data management, facilitates advanced data processing and analysis, and empowers the cooperative to harness the power of data while adhering to its core principles of transparency, accountability, and member empowerment.

\subsection{Social Framework}

\begin{figure}[!h]
\centering
\includegraphics[width=0.6\textwidth]{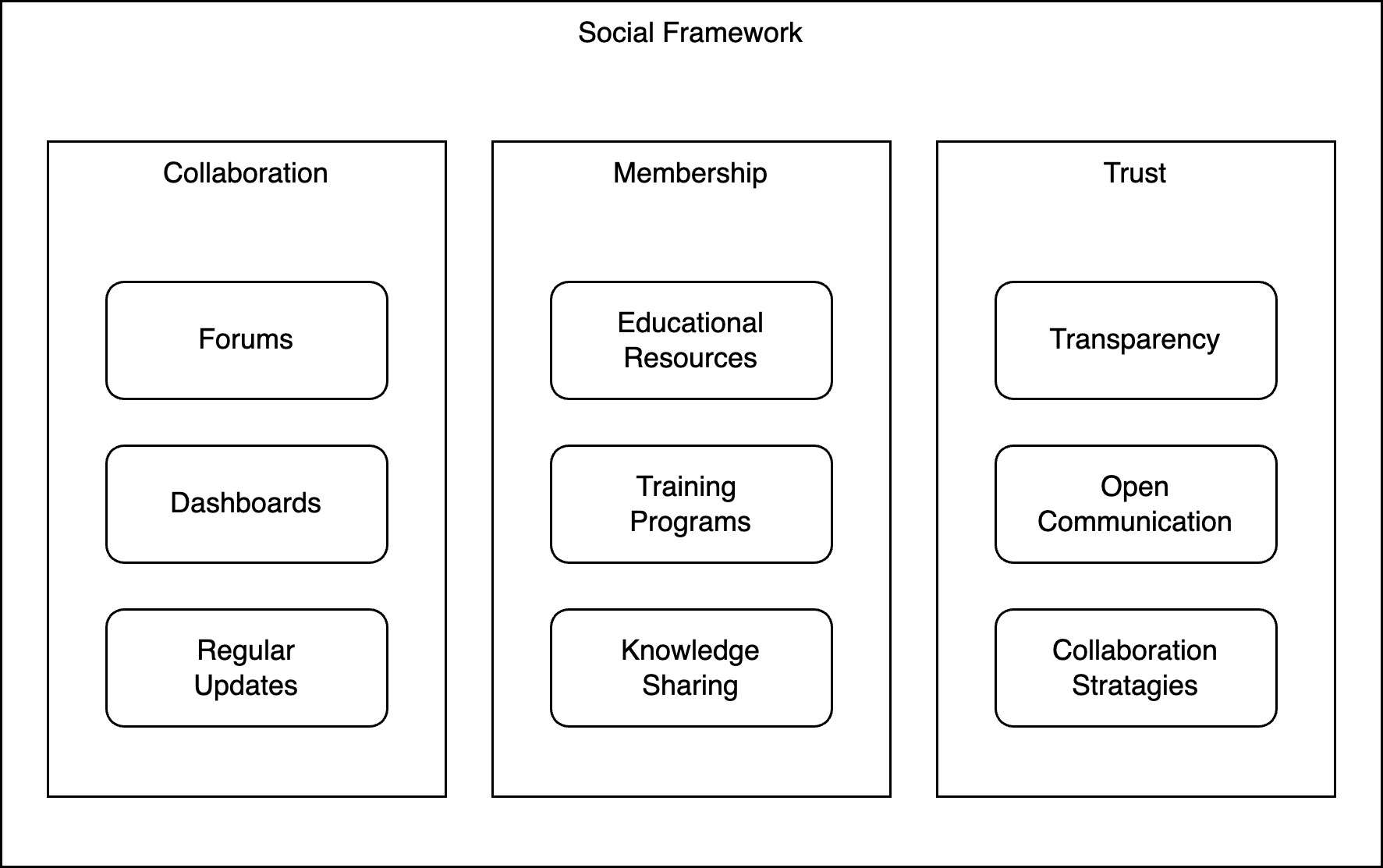}
\caption{Core blocks comprising a Social Framework for a Data Cooperative.}
\label{fig:social-framework}
\end{figure}

Just as with traditional cooperative models, a robust social and community framework (Figure \ref{fig:social-framework}) is indispensable for the success of a data cooperative. This framework goes beyond mere membership; it encompasses deliberate strategies to cultivate member trust, foster active engagement, and encourage collaboration among members, ensuring their meaningful participation in the cooperative's governance and direction.  It recognizes that a data cooperative is not simply a technical system but a community of individuals with shared interests and goals.

This social framework is designed to empower members with the knowledge and tools necessary for effective involvement.  Educational resources, training programs, and workshops can equip members with the understanding of data governance, privacy principles, and the cooperative's operational procedures \cite{Zeuli2005-eu}.  Transparency initiatives are crucial for building trust and keeping members informed about the cooperative's activities and data usage \cite{monitor}.  Regular updates, open forums for discussion, and user-friendly digital dashboards can provide members with clear insights into the cooperative's operations, data management practices, and financial performance.  These initiatives promote open communication and ensure that members are well-informed and can hold the cooperative accountable.

By prioritizing trust and maintaining open communication channels, the cooperative strengthens its community bonds, fostering a sense of shared ownership and purpose.  This strong community foundation is essential for both the integrity of the cooperative's governance and the active involvement of its members.  A thriving community ensures that diverse perspectives are heard, that decisions are made collectively, and that the cooperative remains true to its core principles.

It is crucial to recognize that these four frameworks—governance, operational, technical, and social—are not isolated components.  They are intricately interwoven, each influencing and shaping the others.  They form a dynamic system where decisions made in one framework have repercussions in the others.  For example, decisions made through the governance framework directly influence the operational framework, particularly regarding how data is managed, accessed, and shared.  Member votes or input can shape the cooperative’s data usage policies, which are then implemented by the technical framework through systems that manage access rights, enforce security protocols, and ensure regulatory compliance.

The technical framework, in turn, provides the mechanisms for enforcing the governance decisions. Secure data storage, encryption techniques, and access control protocols are employed to ensure that the cooperative's data is handled responsibly and in accordance with member-approved policies.  Meanwhile, the social and community framework plays a vital role in facilitating member engagement in governance.  By providing members with the necessary tools, education, and fostering a climate of trust, the social framework ensures that members are empowered to actively participate in decision-making processes.  Transparency initiatives, such as dashboards showing data usage statistics or voting results, bridge the gap between governance and operational elements, keeping members informed and enabling them to influence the cooperative’s direction.  These initiatives serve as a feedback loop, allowing members to see the direct impact of their participation and reinforcing their sense of ownership.

The social framework fosters member engagement by equipping individuals with the knowledge, tools, and trust necessary for active participation in governance.  Transparency initiatives, such as interactive dashboards and open communication channels, bridge the gap between governance principles and operational realities, ensuring members remain informed and empowered to shape the cooperative's trajectory.

\begin{figure}[!h]
\centering
\includegraphics[width=0.6\textwidth]{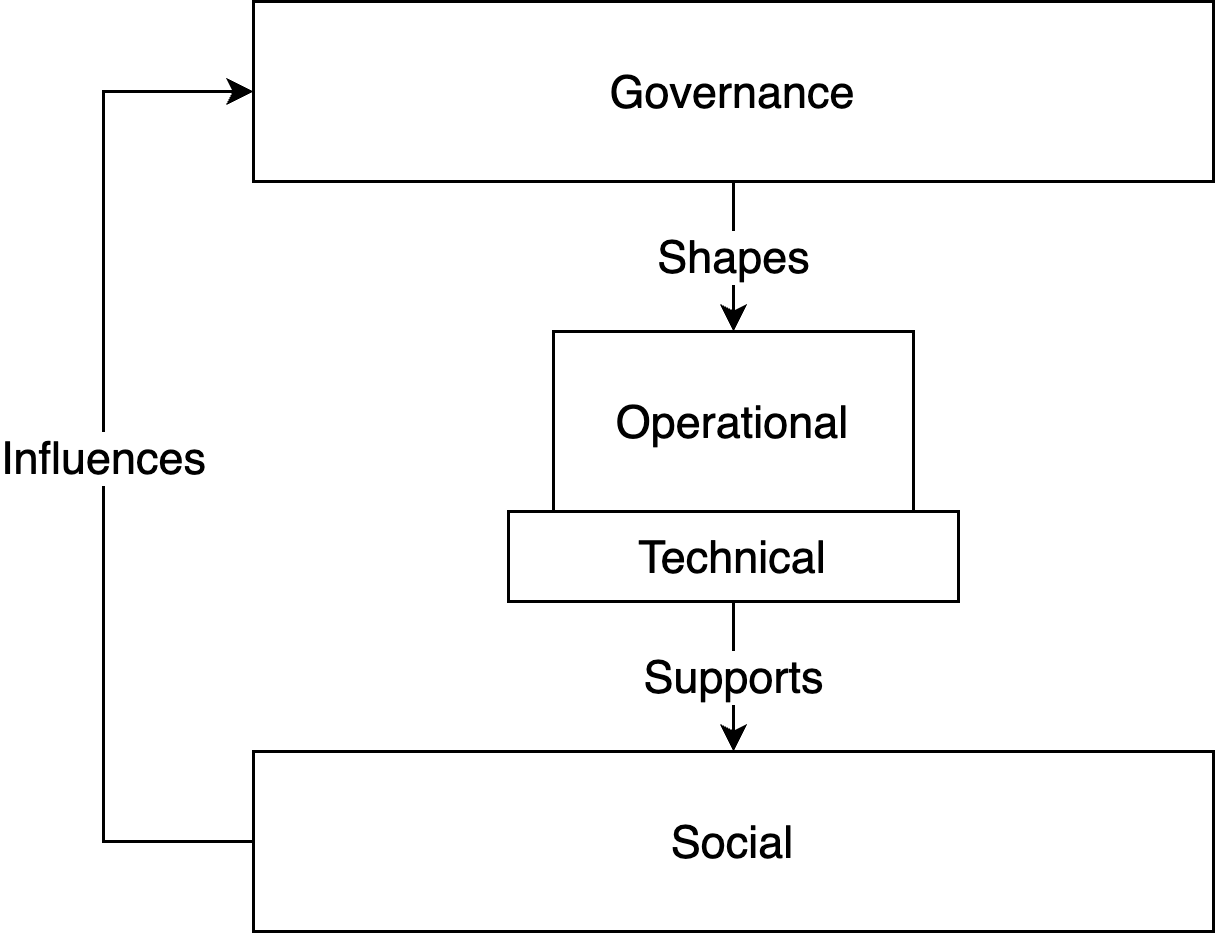}
\caption{Interconnected elements shaping data cooperatives.}
\label{fig:interconnected}
\end{figure}

The four frameworks are not merely interconnected; they exist in a dynamic interplay, forming a continuous feedback loop. Governance decisions shape operational processes, which are in turn supported by the technical infrastructure. Meanwhile, the social framework, by fostering trust and participation, reinforces governance structures, ensuring the cooperative remains accountable to its members and aligned with their collective interests \cite{Wegner2024-jg}. This dynamic interplay highlights the need for a holistic approach, recognizing that each framework is both influenced by and exerts influence upon the others.

\section{Application of Data Cooperatives Across Industries}
\label{applications}

There are several operating data cooperatives across vastly different industries. The most famous is the MIDATA Health Data Cooperative in Switzerland. This cooperative consists of a Data Trust, where users store their data and grant secure access to doctors, researchers, and others based on user-defined rules. \cite{Gille2021-tb}. This is widely cited as a use case for Data Cooperatives, being member-owned, transparent, and secure. They also boast that users can participate in app-based services, thus providing users with “app-based services” and control over the governance of the cooperative through a general assembly \cite{midataHomeMIDATA}.

Some use cases for data cooperatives have been presented in the construction industry. Buhler et al. \cite{Buhler2023-wa} propose using data cooperatives as a compelling opportunity to combat the sector's low productivity and digital maturity by leveraging collaborative data-sharing platforms.

For instance, the construction sector can benefit significantly from pooling data on resource usage, project timelines, and sustainability practices to drive innovation and efficiency \cite{Buhler2023-wa}. One such example is cooperative frameworks to enhance material procurement processes, where shared data allows stakeholders to optimize supply chains and reduce waste, resulting in cost savings and environmental benefits \cite{Buhler2023-vf}. Additionally, integrating digital tools through cooperatives supports more efficient project management, enabling contractors, architects, and SMEs to access real-time information and collaborate seamlessly \cite{Buhler2023-wa}. 

These initiatives demonstrate how cooperatives can transform construction workflows, fostering a data-driven ecosystem that enhances sustainability, reduces project delays, and promotes equitable participation across the industry.

In agriculture, Data cooperatives could be used to offer a robust mechanism to ensure farmers' data sovereignty and promote equitable collaboration, particularly for smallholder farmers. 

For example, initiatives like the Dutch data cooperative JoinData empower farmers to pool and control their data securely while facilitating collaboration with agribusiness and innovation partners \cite{joindataHomepageJoinData}. This model addresses critical challenges such as data lock-in, where farmers are tied to specific technology providers, by enabling independent data management and ensuring farmers retain control over authorizations and data sharing \cite{noauthor_2020-bg}.

Moreover, cooperatives provide a platform for farmers to negotiate better terms for data use and distribution, promoting fair value sharing across the agricultural value chain \cite{Van_der_Burg2021-ub}. 

Data sharing in cooperative frameworks, as seen in models of Cooperative Smart Farming, enhances crop yield optimisation, water resource management, and pest control by leveraging aggregated insights through advanced IoT and machine learning technologies \cite{Gupta2020-mp}. For smallholder farmers, such collective strategies reduce individual costs, democratise access to innovation, and foster resilience by enabling participation in data-driven decision-making, ultimately driving sustainable growth and productivity.

\section{Challenges in Establishing and Scaling Data Cooperatives}
\label{challenges}

Data has emerged as a critical asset in the digital economy, with corporations frequently profiting from personal information in ways that do not always benefit or empower individuals. Data Cooperatives offer a promising alternative by enabling collective ownership and management of data. Rooted in cooperative principles —such as voluntary membership, democratic governance, and community benefit—these cooperatives aim to align data management with the needs and rights of their members.

However, realising the potential of Data Cooperatives presents numerous challenges. From governance complexities to technical and operational hurdles \cite{Susha2017-xr} to social problems, each stage of cooperative growth introduces barriers that complicate ethical data stewardship.

As they scale, cooperatives face additional issues \cite{mehta2022}, including data standardisation, member engagement \cite{aisnetElectronicLibrary}, and maintaining flexible governance structures, all while addressing privacy, security, and ownership concerns inherent to managing intangible assets like data. In this section, we’ll focus on three main points: Coordination, Scalability, and Member-Engagement.

\subsection{Coordination}
Coordination is fundamental to the success of any cooperative, but in Data Cooperatives, the challenges are amplified by the sheer diversity of member backgrounds, data types, and objectives. Managing a large, heterogeneous membership base demands high levels of organisation, transparency, and adaptability, as each member contributes unique perspectives and goals regarding data usage. 

Coordination issues in Data Cooperatives can manifest in three main areas: member decision-making, third-party collaborations, and data standardisation \cite{Susha2017-xr}.

\subsubsection{Member Decision-Making}
Data Cooperatives are inherently democratic, granting members equal voting rights and input on major decisions. However, as membership scales, the cooperative faces a paradox: maintaining democracy while efficiently making decisions. Large groups bring greater diversity but also increase the risk of decision-making deadlock, where divergent interests slow down or halt governance processes \cite{Sacchetto2015-eq}. 

Unlike traditional cooperatives, where members often have shared experiences or sector alignment (e.g., farmers in an agricultural cooperative), Data Cooperatives could comprise members with vastly different expectations about data privacy, monetisation, and governance. This diversity can lead to lengthy negotiations, making it challenging to establish unified policies or adapt to evolving member needs without creating significant delays \cite{opendatamanchesterOpenData}.

\subsubsection{Third-Party Collaboration}
Data Cooperatives often need to engage with third parties—such as researchers, companies, and public institutions—to create value for their members through data sharing, analysis, or product development. Coordinating with these external entities adds complexity, as Data Cooperatives must negotiate terms that protect member interests while achieving mutual benefits \cite{Zhao2021-lu}. 

The cooperative must carefully balance transparency with confidentiality, ensuring members are informed about third-party agreements without compromising sensitive details \cite{Susha2017-xr}. Additionally, negotiating terms that align with cooperative principles, such as fair compensation and ethical data use, requires extensive coordination, as third-party priorities may not always align with the cooperative's mission.

\subsubsection{Data Standardization}
Data Cooperatives combine data from diverse sources, each with its own format, structure, and quality. Without standardisation, the cooperative's ability to pool data and produce valuable insights is severely limited. 

Standardising data across members poses several challenges: differences in data literacy and technical capability among members can lead to inconsistencies in data submission. In contrast, diverse data privacy preferences may require custom data governance rules. For example, some members might be willing to share certain data types (e.g., anonymised usage patterns) but restrict others (e.g., personally identifiable information) \cite{Susha2017-xr}. 

Harmonizing these preferences while ensuring data usability is a significant coordination hurdle, as it requires ongoing efforts to educate members, implement flexible data-sharing policies, and maintain consistent data quality across the cooperative.

Coordination problems within Data Cooperatives present unique challenges that can hinder their growth and effectiveness. Without streamlined processes for member decision-making, third-party engagement, and data standardisation, cooperatives risk losing the trust of their members and the value of collective data. Addressing these issues requires innovative governance structures, clear communication, and tools to automate and facilitate coordination at scale. 

By prioritising these foundational aspects, Data Cooperatives can strengthen their resilience and better serve the collective needs of their members.

\subsection{Scalability}
Scaling a Data Cooperative involves more than just increasing membership or expanding operations; it requires carefully balancing governance, data management, and privacy standards to maintain the cooperative’s core values. Unlike traditional business models, where growth may be managed through centralised decision-making and streamlined operations, Data Cooperatives rely on participatory governance and shared ownership, which introduces unique challenges as they grow.

The complexity of scaling a Data Cooperative can be broken down into three critical areas: governance, technical infrastructure, and cross-sector adaptability.

\subsubsection{Governance Complexity}
Data Cooperatives are governed democratically, which works well in smaller, tightly-knit communities but can become unwieldy as membership grows \cite{Hafen2019-jx}. As cooperatives expand, they must reconcile the cooperative principle of democratic member control with the need for efficient decision-making. Larger memberships mean a greater range of opinions, interests, and expectations, which can slow down governance processes and lead to decision-making gridlock.

Moreover, scaling governance structures may require introducing representative layers—such as elected member councils—to streamline decision-making \cite{Buhler2023-vf}. However, adding such layers can dilute individual member voices and weaken the cooperative’s democratic foundation, risking a drift away from its core cooperative values.

\subsubsection{Technical Infrastructure}
A scalable technical infrastructure is essential for any Data Cooperative to manage and analyse large datasets, enforce privacy standards, and handle increased data traffic \cite{Minzar2024-zb}. 

As membership and data volume grow, cooperatives require robust data storage solutions, secure processing capabilities, and scalable data governance frameworks. However, the decentralised nature of cooperatives often limits their resources, making it challenging to invest in the advanced technologies needed for scaling, such as cloud infrastructure, data security tools, and automated compliance mechanisms \cite{aisnetElectronicLibrary}. 

Moreover, the cooperative must maintain technical flexibility to allow members to control their data access and preferences, which becomes increasingly complex as the scale of the system expands. Without efficient, scalable infrastructure, the cooperative risks data bottlenecks, reduced data quality, and potential security vulnerabilities, all of which could undermine member trust \cite{Buhler2023-vf}.

\subsubsection{Cross-Sector Adaptability}
Scaling a Data Cooperative across different industries introduces further challenges, as each sector may have distinct data standards, regulatory requirements, and governance norms. For instance, a cooperative spanning both healthcare and financial sectors would need to accommodate vastly different data privacy laws, data storage requirements, and member expectations regarding data use \cite{Coche2024-wc}. 

Cross-sector expansion requires adaptability in data governance policies and data-sharing agreements to ensure compliance with sector-specific regulations while upholding cooperative principles \cite{citp}. Additionally, cross-sector cooperatives face complexities in developing a unified business model that balances the distinct value propositions for members from each sector \cite{euiDataAnalytics}. 

These cross-sector complexities often limit scalability, as few data governance frameworks exist that can support cooperative expansion across multiple regulatory-intensive industries.

Scalability remains a central challenge for Data Cooperatives, as expanding membership, technical capabilities, and industry reach can strain the cooperative's foundational structures. Successfully scaling a Data Cooperative requires innovative approaches to governance, investment in flexible and secure technical infrastructure, and adaptive frameworks for cross-sector collaboration. 

By addressing these scalability challenges, Data Cooperatives can broaden their impact and create sustainable, member-centric data management systems that maintain the cooperative’s ethical and democratic values at larger scales

\subsection{Member-Engagement And Trust}
As with all traditional cooperatives, both member trust and participation lie at the heart of the success of any Data Cooperative, as they directly impact both the business efficiency of the cooperative and its ability to draw and retain members.

Unlike conventional data platforms, where users often have limited control or insight into data practices, Data Cooperatives promise transparency, democratic control, and shared benefits \cite{mehta2022}

Challenges in fostering engagement and trust in Data Cooperatives can be understood through three key dimensions: transparency in operations, value alignment, and community-building.

\subsubsection{Transparency in Operations}
For members to engage meaningfully and trust the cooperative, they must clearly understand how their data is used, stored, and shared. Transparent governance requires that members be fully informed about decision-making processes, data usage agreements, third-party collaborations, and any changes in data management policies \cite{Wright2024-os}. However, achieving this level of transparency in a scalable and efficient way is challenging. As Data Cooperatives grow, members may feel disconnected from day-to-day operations or lack access to timely updates on decisions impacting their data \cite{Lauer2024-yl}. 

Building user-friendly dashboards, regular communication channels, and accessible governance tools are essential but can be resource-intensive for cooperatives, particularly when balancing transparency with privacy and operational efficiency. If transparency is not maintained, members may perceive a lack of accountability, which can erode trust and reduce engagement \cite{Joannides_de_Lautour2016-dj}.

\subsubsection{Value Alignment and Tangible Benefits}
Maintaining engagement requires more than just transparency; members must see tangible benefits that align with their values and expectations. A core challenge for Data Cooperatives is balancing the collective interests of the membership with individual members’ diverse motivations, which may include data protection, monetization, or ethical data use \cite{mehta2022}. 

Providing clear, individualized benefits—such as dividends from data-sharing profits, access to valuable insights, or exclusive services—can help strengthen members' commitment to the cooperative \cite{euiDataAnalytics}. However, designing a benefit structure that meets varied needs without compromising cooperative principles can be complex, particularly as the cooperative scales and member priorities diversify. Without perceivable value, members may disengage, reducing the cooperative’s active community and weakening its democratic foundation.

\subsubsection{Community-Building and Engagement Strategies}
Fostering a strong sense of community is vital for keeping members engaged and reinforcing the cooperative’s identity. Data Cooperatives differ from traditional digital platforms in that they are intended to function as member-centric communities rather than service providers \cite{data2x}. However, such a sense of community can be hard to create, especially in online communities where members are geographically dispersed or lack interpersonal connections. Regular community events, discussion forums, and decision-making workshops can make members feel more connected to the cooperative and more invested in its success \cite{oecd}. Giving members mechanisms to participate in governance—such as voting on key issues, serving on advisory committees, or leading projects—can also enhance members' sense of agency and commitment. However, these initiatives require resources and strategic planning to sustain, especially as the cooperative grows, increasing the administrative burden and the complexity of integrating member input.

Building and maintaining member engagement and trust is essential but challenging for Data Cooperatives. Without trust and active participation, a Data Cooperative risks losing its ethical and democratic edge, which distinguishes it from traditional data platforms \cite{Buang2021-mf}. Addressing these challenges requires investing in transparent governance \cite{Bauer2019-oh}, aligning value with member expectations, and fostering a strong sense of community through intentional engagement strategies. By prioritising these aspects, Data Cooperatives can create a robust foundation of trust and engagement that supports long-term sustainability and success.

\section{Future Research}
This paper has laid the groundwork for understanding the potential of data cooperatives. However, translating this potential into reality requires addressing key open questions, particularly when concerning cross-industry collaboration and the practical implementation of robust governance frameworks.  Future research should delve into the following crucial areas:

A significant challenge lies in designing data governance frameworks that facilitate cooperation and can be applied to diverse industries.  Current models often focus on single-sector applications.  However, many societal challenges (e.g., sustainable development, public health) require data sharing and collaboration across sectors. For example, a valid use case would be the optimization of value chains. Sharing data within a cooperative framework has the potential to create a more holistic and integrated system, leading to efficiencies and improvements throughout the entire value chain. Future investigations should explore:

\begin{itemize}
    \item How to harmonize disparate data standards, regulatory requirements, and ethical norms, which are prevalent across different industries, within a unified data cooperative structure? This might be achieved the modularization of the data cooperative structure, and will include tackling legal interoperability issues.
    \item Which governance mechanisms can effectively manage the intricate power dynamics and potentially conflicting interests that may arise when multiple industries participate in a shared data cooperative? This may involve the exploration of different voting systems, the possible integration of industry-specific councils within the cooperative, or other tailored governance structures.
    \item How to create robust incentive structure that encourage member participation while simultaneously ensuring equitable distribution of benefits and safeguarding sensitive competitive information? This could involve exploring tiered membership models, differential data access structures, or the development of secure data enclaves for particularly sensitive information. 
\end{itemize}

In short, the future of data cooperatives rests on solving key challenges such as harmonizing cross-industry data governance, optimizing value chains through data sharing, and building the necessary technological and operational infrastructure. Future research must prioritize modular governance design approaches, robust incentive mechanisms, and effective data analytics in order to unlock the full potential of data cooperatives for innovation and societal benefit.

\section{Conclusion}
Data cooperatives are a new approach to data governance in which individuals and organisations can jointly own and benefit from their data resources. By adhering to cooperative values—among which are democratic member control, economic participation, and concern for the community—these organisations offer a viable alternative to traditional models of data management by big corporations.

The establishment of data cooperatives, while still in their initial phases, in various sectors, including healthcare, agriculture, and construction, demonstrates their potential in addressing industry-specific challenges by collectively sharing and governing data. For example, agricultural data cooperatives can empower farmers to negotiate better terms for data use and dissemination, promoting fair value distribution across the agricultural value chain.

Yet, constructing and developing data cooperatives is extremely demanding. Cooperation among heterogeneous members, scalability of governance and technical infrastructures, and long-term participation of members are essential dimensions that shape the success of these cooperatives. Addressing these challenges requires innovative governance designs, robust technical solutions, and institutions for establishing trust and active member engagement.

In conclusion, data cooperatives are a promising model for equitable data governance, enabling communities to retain the value of their collective data while upholding democratic principles and individual rights. By leveraging existing implementations and addressing inherent complexities, data cooperatives can play a vital role in shaping the data economy into an inclusive and participatory one. A key area of future research is the study of cross-industry collaboration models and the development of scalable, member-led governance models to realize this potential to its fullest.

\bibliographystyle{unsrt}

\end{document}